\documentclass{elsarticle}
\usepackage{graphicx}
\usepackage{amsmath}
\usepackage{amsfonts}
\usepackage{amssymb}
\usepackage{dirtree}
\usepackage{subfigure}

\usepackage{listings}
\newcommand{\belowtitleskip}{2pt}

\usepackage{xcolor}
\definecolor{gray}{gray}{0.97}
\colorlet{commentcolour}{green!50!black}
\colorlet{stringcolour}{red!60!black}
\colorlet{keywordcolour}{magenta!90!black}
\colorlet{exceptioncolour}{yellow!50!red}
\colorlet{commandcolour}{blue!60!black}
\colorlet{promptcolour}{green!50!black}

\lstdefinestyle{pythonstyle}{
keepspaces=true,
language=python,
showtabs=true,
tab=,
tabsize=2,
basicstyle=\ttfamily\footnotesize,
stringstyle=\color{stringcolour},
showstringspaces=false,
otherkeywords={\ , \}, \{, \%, \&, \|},
keywordstyle=\color{keywordcolour}\bfseries,%
emph={[2]True, False, None},
emphstyle=[2]\color{keywordcolour},
emph={[3]object,type,isinstance,deepcopy,zip,enumerate,
reversed,list,len,dict,tuple,xrange,append,execfile,real,imag,
reduce,str,repr, vars},
emphstyle=[3]\color{commandcolour},
emph={Exception,NameError,IndexError,SyntaxError,TypeError,ValueError,OverflowError,ZeroDivisionError},
emphstyle=\color{exceptioncolour}\bfseries,
morestring=[s]{"""}{"""},
morestring=[s]{'''}{'''},
commentstyle=\color{commentcolour}\slshape,
emph={[4]1, 2, 3, 4, 5, 6, 7, 8, 9, 0, 001, ode, fsolve, sqrt, exp, sin, cos, arccos, array, norm, dot, arange, , isscalar, max, sum, flatten, reshape, find, any, all, abs, plot, linspace, legend, quad, polyval,polyfit, hstack, concatenate,vstack,column_stack,empty,rand,vander,
grid,pcolor,eig,eigs,eigvals,svd,qr,tan,det,logspace,roll,min,
mean,cumsum,cumprod,vectorize,lstsq,cla,eye,xlabel,ylabel,
},
emphstyle=[4]\color{commandcolour},
emph={[5]and,break,class,continue,def,yield,del,elif ,else,
except,exec,finally,for,global,if,in,
lambda,not,or,pass,print,raise,return,try,while,assert},
emphstyle=[5]\color{black}\bf,
emph={[6]Function,TestFunction,TrialFunction,TensorProductSpace,C2CBasis,FourierBasis,R2CBasis,ShenDirichletBasis,ShenBiharmonicBasis,Basis,VectorTensorProductSpace},
emphstyle=[6]\color{black},
emph={[7]inner,project,div,grad,Dx,curl},
emphstyle=[7]\color{black},
literate=*%
{>>>}{{\textcolor{promptcolour}{>>>}}}{1}%
,%
breaklines=true,
breakatwhitespace= true,
xleftmargin=0ex,
xrightmargin=0ex,
aboveskip=1ex,
belowskip=1ex,
frame=trbl, 
numbers=none,
backgroundcolor=\color{gray}
}

\lstdefinestyle{inlinestyle}{
language=python,
basicstyle=\ttfamily,
breaklines=true,
breakatwhitespace= true,
xleftmargin=0ex,
xrightmargin=0ex,
aboveskip=1ex,
belowskip=1ex,
frame=trbl, 
numbers=none,
float=htpb,
}

\lstdefinestyle{scriptstyle}{
language=python,
numbers=left,
}

\lstnewenvironment{python}[1][]{
\lstset{style=pythonstyle, frame=trbl, belowcaptionskip=\belowtitleskip}
}{}

\lstnewenvironment{python_num}[1][]{
\lstset{style=pythonstyle, numbers=left, frame=trbl, 
belowcaptionskip=\belowtitleskip}
}{}

\newcommand{\inpyth}{\lstinline[style=inlinestyle]} %[]%

\newcommand{\bs}[1]{\boldsymbol{#1}}
\newcommand{\ts}[1]{\bs{\textsf{#1}}}

\begin{document}
\begin{frontmatter}
\title{Shenfun - automating the spectral Galerkin method}

\author[mmo]{Mikael Mortensen}
\ead{mikaem@math.uio.no}
\address[mmo]{Department of Mathematics, Division of Mechanics, University of Oslo}


\begin{abstract}{With the \emph{shenfun} Python module (github.com/spectralDNS/shenfun) an effort is made towards automating the implementation of the spectral Galerkin method for simple tensor product domains, consisting of (currently) one non-periodic and any number of periodic directions. The user interface to \emph{shenfun} is intentionally made very similar to FEniCS (fenicsproject.org). Partial Differential Equations are represented through weak variational forms and solved using efficient direct solvers where available. MPI decomposition is achieved through the {mpi4py-fft} module (bitbucket.org/mpi4py/mpi4py-fft),  and all developed solver may, with no additional effort, be run on supercomputers using thousands of processors. Complete solvers are shown for the linear Poisson and biharmonic problems, as well as the nonlinear and time-dependent Ginzburg-Landau equation. }
\end{abstract}
\end{frontmatter}
\section{INTRODUCTION}

The spectral Galerkin method, see, e.g., Shen \cite{shen95} or Kopriva \cite{kopriva09}, combines spectral basis functions with the Galerkin method and allows for highly accurate solutions on simple, tensor product domains. Due to its accuracy and efficiency, the method is often favoured in studies of sensitive fundamental physical phenomena, where numerical errors needs to be avoided. 

In this paper we will describe the \inpyth{shenfun} Python module. The purpose of \inpyth{shenfun} is to simplify the implementation of the spectral Galerkin method, to make it easily accessible to researchers, and to make it easier to solve advanced PDEs on supercomputers, with MPI, in simple tensor product domains. The package can solve equations for tensor product spaces consisting of any number of periodic directions, but, at the moment of writing, only one non-periodic direction. This configuration may sound trivial, but it occurs surprisingly often in physics, for example in plane shear flows like the channel or pipe. And these simple configurations are used heavily to enhance our understanding of fundamental physical processes, like turbulence, or transition to turbulence, turbulent mixing, and turbulent combustion.

The \inpyth{shenfun} package is heavily influenced by the FEniCS project \cite{fenics}, that has made it trivial to solve PDEs in arbitrary complex domains with the finite element method (FEM). FEM also makes use of the Galerin method to set up variational forms. However, where FEM uses basis functions with only local support, the spectral Galerkin method uses basis functions with global support. The local support is one of the many nice features of the FEM, which makes it particularly attractive for unstructured and complex geometries. Spectral methods, on the other hand, are less flexible, but represent the gems of numerical methods, and, whenever possible, when the domain is simple and the solution is smooth, delivers the most accurate approximations.

There are many tools available for working with spectral methods. For MATLAB there is the elegant chebfun package \cite{trefethen13}, with an extensive list of application for, e.g., PDEs, ODEs or eigenvalue problems. However, being implemented in MATLAB, there is no feasible extension to DNS and supercomputers through MPI. Numpy and Scipy have modules for orthogonal polynomials (Jacobi, Chebyshev, Legendre, Hermite), and for Fourier transforms, which are both utilized by \inpyth{shenfun}. The orthogonal module makes it easier to work with Chebyshev and Legendre polynomials, as it delivers, for example, quadrature points and weights for different quadrature rules (e.g., Chebyshev-Gauss, Legendre-Gauss). 

To the author's knowledge, all research codes developed for studying turbulent flows through Direct Numerical Simulations (DNS) on supercomputers have been written in low-level languages like Fortran, C or C++, see, e.g., \cite{debruynkops15,hoyas06,leemoser15}, or \cite{Alfonsi2016} for a list of high performance channel flow solvers. The codes are  highly tuned and tailored to a specific target, and, being low-level, the codes are not easily accessible to a non-expert programmer. Mortensen and Langtangen \cite{Mortensen2016} describe how a DNS solver can be written in Python in 100 lines of script-like code, and also show that the code, when optimized in the background using Cython, runs as fast as an identical C++ implementation on thousands of processors with MPI. \inpyth{Shenfun} takes it one step further and aims at providing a generic toolbox for creating high performance, parallel solvers of any PDE, in a very high-level language. And without compromising much on computational efficiency. The key to developing such a high-level code in Python is efficient use of Numpy \cite{numpy}, with broadcasting and vectorization, and MPI for Python \cite{mpi4py}, that wraps almost the entire MPI library, and that can transfer Numpy arrays between thousands of processors at the same speed as a low-level C or Fortran code. Similarly, we utilize the pyFFTW module \cite{pyfftw}, that wraps most of the FFTW library \cite{libfftw} and makes the FFT as fast when called from Python as it is when used in low-level codes.

This paper is organised as follows: in Sec \ref{sec:preliminaries} the spectral Galerkin method is introduced. In Sec.~\ref{sec:shenfun} the basics of the \inpyth{shenfun} package is described and implementations are shown for simple 1D Poisson and biharmonic problems. In Sec~\ref{sec:tensorproductspaces} we move to higher dimensions and tensor product spaces before we, in Secs~\ref{sec:extended} and \ref{sec:ginzburg} end with some extended functionality and an implementation for the time dependent nonlinear Ginzburg-Landau equation in 2D.

\section{SPECTRAL GALERKIN METHOD}
\label{sec:preliminaries}
The spectral Galerkin method can most easily be described by considering a simple PDE, like the Poisson equation, in a 1D domain $\Omega$
\begin{equation}
-u''(x) = f(x), \quad x \in \Omega, \label{eq:poisson}
\end{equation}
with appropriate boundary conditions (Dirichlet, Neumann or periodic). To solve this equation, we can define a test function $v(x)$ that satisfies the boundary conditions, and that comes with an accompanying weight function $w(x)$. Assuming also that we work with complex valued functions, a weighted continuous inner product of the two functions $u$ and $v$ can be defined as
\begin{equation}
(u, v)_w = \int_{\Omega} u(x) \overline{v}(x) w(x) dx,
\end{equation}
where $\overline{v}$ is the complex conjugate of $v$. The weighted inner product can now be used to create variational forms. If we multiply Eq.~(\ref{eq:poisson}) with $\overline{v}w$ and integrate over the domain we obtain the variational form of the PDE
\begin{equation}
(-u'', v)_w = (f, v)_w.
\label{eq:weak_poisson}
\end{equation}
The variational form can be solved numerically if $u$ and $v$ are approximated using a finite number $(N)$ of test functions $\{v_l(x)\}_{l=0}^{N-1}$, and a solution 
\begin{equation}
u(x) = \sum_{l=0}^{N-1} \hat{u}_l v_l(x),
\end{equation}
where $\bs{\hat{u}} = \{\hat{u}_l\}_{l=0}^{N-1}$ are the expansion coefficients, that are also recognised as the unknowns in the modal spectral Galerkin method.

If $v$ is chosen from a Fourier or Legendre basis, then the weight function used in the inner product is simply constant, and we may integrate (\ref{eq:weak_poisson}) further using integration by parts. However, for a Chebyshev basis the weight function will be $1/\sqrt{1-x^2}$ and integration by parts is thus usually avoided. The weighted continuous inner product may, depending on the function that is to be integrated, be difficult or costly to evaluate. As such, we will in this work use the weighted \emph{discrete} inner product instead, where the integral is approximated using quadrature
\begin{equation}
(u, v)_w^N = \sum_{j=0}^{N-1} u(x_j) \overline{v}(x_j) w_j  \approx  \int_{\Omega} u(x) \overline{v}(x) w(x) dx.
\label{eq:quadrature}
\end{equation}
Here $\{w_j\}_{j=0}^{N-1}$ represents the quadrature weights and $\{x_j\}_{j=0}^{N-1}$ are the quadrature points for the integration. 

The test functions $v$ will be chosen based in part on boundary conditions. However, regardless of which space the test functions are chosen from, the procedure for solving a PDE with the spectral Galerkin method is always the same:
\begin{enumerate}
\item Choose a basis satisfying boundary conditions.
\item Derive variational forms from PDEs using  weighted inner products.
\item Assemble and solve linear systems of equations for expansion coefficients.
\end{enumerate}
In other words it is very much like a finite element method. The major difference is that the basis functions are global, i.e., they all span the entire domain, whereas in FEM the test functions only have local support.

\section{SHENFUN}
\label{sec:shenfun}
\inpyth{Shenfun} is a Python module package containing tools for working with the spectral Galerkin method. Shenfun implements classes for several bases with different boundary conditions, and within each class there are methods for transforms between spectral and real space, inner products, and for computing matrices arising from bilinear forms in the spectral Galerkin method. The Python module is organized as shown in Figure~\ref{fig:directorytree}. 

The \inpyth{shenfun} language is very simple and closely follows that of FEniCS. A simple form implementation provides operators \inpyth{div, grad, curl} and \inpyth{Dx}, that act on three different types of basis functions, the \inpyth{TestFunction, TrialFunction} and \inpyth{Function}. Their usage is very similar to that from FEniCS, but not as general, nor flexible, since we are only conserned with simple tensor product grids and smooth solutions. The usage of these operators and basis functions will become clear in the following subchapters, where we will also describe the \inpyth{inner} and \inpyth{project} functions, with functionality as suggested by their names.
\begin{figure}[h!]
\dirtree{%
 .1 shenfun. 
 .2 {\_\_init\_\_.py}.
 .2 {spectralbase.py}.
 .2 {matrixbase.py}.
 .2 {tensorproductspace.py}.
 .2 {la.py}. 
 .2 chebyshev/.
 .3 {\_\_init\_\_.py}.
 .3 {bases.py}.
 .3 {matrices.py}.
 .3 {la.py}.
 .2 legendre/.
 .3 {\_\_init\_\_.py}.
 .3 {bases.py}.
 .3 {matrices.py}.
 .3 {la.py}.
 .2 fourier/.
 .3 {\_\_init\_\_.py}.
 .3 {bases.py}.
 .3 {matrices.py}.
 .2 forms/.
 .3 {operators.py}.
 .3 {inner.py}.
 .3 {arguments.py}.
}
\caption{Directory tree structure of Python package \inpyth{shenfun}. }
\label{fig:directorytree}
\end{figure}

\subsection{Classes for basis functions}
The following bases are defined in submodules
\begin{itemize}
\item \inpyth{shenfun.chebyshev.bases}
  \begin{itemize}
    \item \inpyth{Basis} - Regular Chebyshev 
    \item \inpyth{ShenDirichletBasis} - Dirichlet boundary conditions
    \item \inpyth{ShenNeumannBasis} - Neumann boundary conditions (homogeneous)
    \item \inpyth{ShenBiharmonicBasis} - Homogeneous Dirichlet and Neumann boundary conditions
  \end{itemize}
\item \inpyth{shenfun.legendre.bases}
  \begin{itemize}
    \item \inpyth{Basis} - Regular Legendre
    \item \inpyth{ShenDirichletBasis} - Dirichlet boundary conditions
    \item \inpyth{ShenNeumannBasis} - Neumann boundary conditions (homogeneous)
    \item \inpyth{ShenBiharmonicBasis} - Homogeneous Dirichlet and Neumann boundary conditions
  \end{itemize}

 \item \inpyth{shenfun.fourier.bases}
  \begin{itemize}
    \item \inpyth{R2CBasis} - Real to complex Fourier transforms
    \item \inpyth{C2CBasis} - Complex to complex transforms
  \end{itemize}
\end{itemize}

All bases have methods for transforms and inner products on single- or multidimensional Numpy data arrays. The following code shows how to create a Fourier basis and subsequently perform a forward and an inverse discrete Fourier transform on a random array. The \inpyth{uc} array is only used to test that the transform cycle returns the original data.
\vskip 1ex
\noindent
\begin{minipage}{\columnwidth}
\begin{python}
    >>> from shenfun import *
    >>> import numpy as np
    >>> N = 16
    >>> FFT = fourier.bases.R2CBasis(N, plan=True) 
    >>> u = np.random.random(N)
    >>> uc = u.copy()
    >>> u_hat = FFT.forward(u)
    >>> u = FFT.backward(u_hat) 
    >>> assert np.allclose(u, uc)
\end{python}
\end{minipage}

\subsection{Classes for matrices}
Matrices that arise with the spectral Galerkin method using Fourier or Shen's modified basis functions (see, e.g., Eqs~(\ref{eq:chebdirichlet}, \ref{eq:legdirichlet})), are typically sparse and diagonal in structure. The sparse structure allows for a very compact storage, and \inpyth{shenfun} has its own Matrix-class that is subclassing a Python dictionary, where keys are diagonal offsets, and values are the values along the diagonal. Some of the more important methods of the \inpyth{SparseMatrix} class are shown below:
\vskip 1ex
\noindent
\begin{minipage}{\columnwidth}
\begin{python}
class SparseMatrix(dict):
    def __init__(self, d, shape):
        dict.__init__(self, d)
        self.shape = shape
        
    def diags(self, format='dia'):
        """Return Scipy sparse matrix"""

    def matvec(self, u, x, format='dia', axis=0):
        """Return Matrix vector product self*u in x"""
        
    def solve(self, b, u=None, axis=0):
        """Return solution u to self*u = b"""

\end{python}
\end{minipage}
For example, we may declare a tridiagonal matrix of shape N x N as
\vskip 1ex
\noindent
\begin{minipage}{\columnwidth}
\begin{python}
    >>> N = 4
    >>> d = {-1: 1, 0: -2, 1: 1}
    >>> A = SparseMatrix(d, (N, N))
\end{python}
\end{minipage}
or similarly as
\vskip 1ex
\noindent
\begin{minipage}{\columnwidth}
\begin{python}
    >>> d = {-1: np.ones(N-1), 0: -2*np.ones(N)}
    >>> d[1] = d[-1]  # Symmetric, reuse np.ones array
    >>> A = SparseMatrix(d, (N, N))
    >>> A
    {-1: array([ 1.,  1.,  1.]),
      0: array([-2., -2., -2., -2.]),
      1: array([ 1.,  1.,  1.])}
\end{python}
\end{minipage}
The matrix is a subclassed dictionary. If you want a regular \inpyth{Scipy} sparse matrix instead, with all of its associated methods (solve, matrix-vector, etc.), then it is just a matter of
\vskip 1ex
\noindent
\begin{minipage}{\columnwidth}
\begin{python}
    >>> A.diags()
    <4x4 sparse matrix of type '<class 'numpy.float64'>'
        with 10 stored elements (3 diagonals) in DIAgonal format>
    >>> A.diags().toarray()
    array([[-2.,  1.,  0.,  0.],
           [ 1., -2.,  1.,  0.],
           [ 0.,  1., -2.,  1.],
           [ 0.,  0.,  1., -2.]])
\end{python}
\end{minipage}
\subsection{Variational forms in 1D}
Weak variational forms are created using test and trial functions, as shown in Sec~\ref{sec:preliminaries}. Test and trial functions can be created for any basis in \inpyth{shenfun}, as shown below for a Chebyshev Dirichlet basis with 8 quadrature points
\vskip 1ex
\noindent
\begin{minipage}{\columnwidth}
\begin{python}
    >>> from shenfun.chebyshev.bases import ShenDirichletBasis
    >>> from shenfun import inner, TestFunction, TrialFunction    
    >>> N = 8
    >>> SD = ShenDirichletBasis(N, plan=True)
    >>> u = TrialFunction(SD)
    >>> v = TestFunction(SD)
\end{python}
\end{minipage}
A matrix that is the result of a bilinear form has its own subclass of \inpyth{SparseMatrix}, called a \inpyth{SpectralMatrix}. A \inpyth{SpectralMatrix} is created using \inpyth{inner} products on test and trial functions, for example the mass matrix:
\vskip 1ex
\noindent
\begin{minipage}{\columnwidth}
\begin{python}
    >>> mass = inner(u, v)
    >>> mass
    {-2: array([-1.57079633]),
      0: array([ 4.71238898,  3.1415
                 3.14159265, 3.14159265]),
      2: array([-1.57079633])}
\end{python}
\end{minipage}
This \inpyth{mass} matrix will be the same as Eq. (2.5) of \cite{shen95}, and it will be an instance of the \inpyth{SpectralMatrix} class.
You may notice that \inpyth{mass} takes advantage of the fact that two diagonals are constant and consequently only stores one single value.

The \inpyth{inner} method may be used to compute any linear or bilinear form. For example the stiffness matrix \inpyth{K}
\vskip 1ex
\noindent
\begin{minipage}{\columnwidth}
\begin{python}
    >>> K = inner(v, div(grad(u)))
\end{python}
\end{minipage}

Square matrices have implemented a solve method that is using fast $\mathcal{O}(N)$ direct LU decomposition or similar, if available, and falls back on using Scipy's solver in CSR format if no better method is found implemented. For example, to solve the linear system \inpyth{Ku=b}
\vskip 1ex
\noindent
\begin{minipage}{\columnwidth}
\begin{python}
    >>> fj = np.random.random(N)
    >>> b = inner(v, fj)
    >>> u = np.zeros_like(b)
    >>> u = K.solve(b, u)
\end{python}
\end{minipage}
All methods are designed to work along any dimension of a multidimensional array. Very little differs in the users interface. Consider, for example, the previous example on a three-dimensional cube 
\begin{python}
    >>> fj = np.random.random((N, N, N))
    >>> b = inner(v, fj)
    >>> u = np.zeros_like(b)
    >>> u = K.solve(b, u)
\end{python}
where \inpyth{K} is exactly the same as before, from the 1D example. The matrix solve is applied along the first dimension since this is the default behaviour.

The bases also have methods for transforming between spectral and real space. For example, one may project a random vector to the \inpyth{SD} space using
\vskip 1ex
\noindent
\begin{minipage}{\columnwidth}
\begin{python}
    >>> fj = np.random.random(N)
    >>> fk = np.zeros_like(fj)
    >>> fk = SD.forward(fj, fk) # Gets expansion coefficients 
\end{python}
\end{minipage}
and back to real physical space again
\begin{python}
    >>> fj = SD.backward(fk, fj)
\end{python} 
Note that \inpyth{fj} now will be different than the original \inpyth{fj} since it now has homogeneous boundary conditions. However, if we transfer back and forth one more time, starting from \inpyth{fj} which is in the Dirichlet function space, then we come back to the same array:
\begin{python}
    >>> fj_copy = fj.copy()
    >>> fk = SD.forward(fj, fk)
    >>> fj = SD.backward(fk, fj)
    >>> assert np.allclose(fj, fj_copy) # Is True
\end{python}

\subsection{Poisson equation implemented in 1D}
We have now shown the usage of \inpyth{shenfun} for single, one-dimensional spaces. It does not become really interesting before we start looking into tensor product grids in higher dimensions, but before we go there we revisit the spectral Galerkin method for a 1D Poisson problem, and show how the implementation of this problem can be performed using \inpyth{shenfun}.

\subsubsection{Periodic boundary conditions}
\label{sec:fourierpoisson}
If the solution to Eq.~(\ref{eq:poisson}) is periodic with periodic length $2 \pi$, then we use $\Omega \in [0, 2 \pi]$ and it will be natural to choose the test functions from the space consisting of the Fourier basis functions, i.e.,  $v_l(x)=e^{ilx}$. The mesh $\boldsymbol{x} = \{x_j\}_{j=0}^{N-1}$ will be uniformly spaced 
\begin{equation}
\boldsymbol{x} = \frac{2 \pi j}{N}  \quad j=0,1,\ldots, N-1,
\end{equation}
and we look for solutions of the form
\begin{equation}
u(x_j) = \sum_{l=-N/2}^{N/2-1} \hat{u}_l e^{ilx_j} \quad  j=0,1,\ldots N-1.
\label{eq:ufourier}
\end{equation}
Note that for Fourier basis functions it is customary (used by both MATLAB and Numpy) to use the wavenumbermesh
\begin{equation}
\boldsymbol{l} = -N/2, -N/2+1, \ldots, N/2-1, \label{eq:wavenumber_even}
\end{equation}
where we have assumed that $N$ is even. Also note that Eq.~(\ref{eq:ufourier}) naively would be computed in $\mathcal{O}(N^2)$ operations, but that it can be computed much faster $\mathcal{O}(N\log N)$ using the discrete inverse Fourier transform
\begin{equation}
\bs{u} = \mathcal{F}^{-1}(\bs{\hat{u}}),
\end{equation}
where we use compact notation $\bs{u} = \{u(x_j)\}_{j=0}^{N-1}$.

To solve Eq.(\ref{eq:poisson}) with the discrete spectral Galerkin method, we create the basis $V^p = \text{span}\{ e^{ilx} , \text{ for } l \in \boldsymbol{l}\} $ and attempt to find $u \in V^p$ such that
\begin{equation}
(-u'', v)_w^N = (f, v)_w^N, \quad \forall \, v \in V^p.
\end{equation}
Inserting for Eq. (\ref{eq:ufourier}) and using $e^{imx}$ as test function we obtain
\begin{align}
-(\sum_{l \in \bs{l}} \hat{u}_l (e^{ilx})'', e^{imx})_w^N &= (f(x), e^{imx})_w^N \quad \forall \, m \in \bs{l} \\
\sum_{l \in \bs{l}} l^2( e^{ilx}, e^{imx})_w^N \hat{u}_l &= (f(x), e^{imx})_w ^N\quad \forall \, m \in \bs{l}. \label{eq:utmp}
\end{align}
Note that the discrete inner product (\ref{eq:quadrature}) is used, and we also need to interpolate the function $f(x)$ onto the grid $\boldsymbol{x}$. For Fourier it becomes very simple since the weight functions are constant $w_j = 2\pi/N$ and we have for the left hand side simply a diagonal matrix
\begin{equation}
( e^{ilx}, e^{imx})^N = 2\pi \delta_{ml} \quad \text{for} \, l, m \in \bs{l} \times \bs{l},
\end{equation}
where $\delta_{ml}$ is the kronecker delta function.
For the right hand side we have
\begin{align}
(f(x), e^{imx})^N &= \frac{2 \pi}{N}\sum_{j=0}^{N-1} f(x_j) e^{-imx_j} \quad \text{for } m \in \bs{l}, \\
 &= 2 \pi \mathcal{F}_m(f(\bs{x})), \\
 &= 2 \pi \hat{f}_m,
\end{align}
where $\mathcal{F}$ represents the discrete Fourier transform that is defined as
\begin{equation}
\hat{u}_l = \frac{1}{N}\sum_{j=0}^{N-1} u(x_j) e^{-ilx_j}, \quad \text{for } l \in \bs{l},
\end{equation}
or simply
\begin{equation}
  \bs{\hat{u}} = \mathcal{F}(\bs{u}).
\end{equation}
Putting it all together we can set up the assembled linear system of equations for $\hat{u}_l$ in (\ref{eq:utmp})
\begin{equation}
\sum_{l \in \bs{l}}2 \pi l^2 \delta_{ml} \hat{u}_l = 2 \pi \hat{f}_{m} \quad \forall \, m \in \bs{l},
\end{equation}
which is trivially solved since it only involves a diagonal matrix ($\delta_{ml}$), and we obtain
\begin{equation}
\hat{u}_l = \frac{1}{l^2} \hat{f}_{l} \quad \forall \,l  \in \bs{l} \setminus{\{0\}}.
\end{equation}
 
So, even though we carefully followed the spectral Galerkin method, we have ended up with the same result that would have been obtained with a Fourier collocation method, where one simply takes the Fourier transform of the Poisson equation and differentiate analytically.

With \inpyth{shenfun} the periodic 1D Poisson equation can be trivially computed either with the collocation approach or the spectral Galerkin method. The procedure for the spectral Galerkin method will be shown first, and in Figure \ref{fig:poisson1D}, the entire problem is solved.
 All \inpyth{shenfun} demos in this paper will contain a similar preample section where some necessary Python classes, modules and functions are imported. We import Numpy since \inpyth{shenfun} arrays are Numpy arrays, and we import from Sympy to construct some exact solution used to verify the code. Note also the similarity to FEniCS with the import of methods and classes \inpyth{inner, div, grad, TestFunction, TrialFunction}.  The Fourier spectral Galerkin method in turn requires that the \inpyth{FourierBasis} is imported as well. 
\begin{figure}[h!]
\begin{python}
from sympy import Symbol, cos
import numpy as np
from shenfun import inner, div, grad, TestFunction, TrialFunction
from shenfun.fourier.bases import FourierBasis

# Use Sympy to compute a rhs, given an analytical solution
x = Symbol("x")
ue = cos(4*x)
fe = ue.diff(x, 2)

# Create Fourier basis with N basis functions
N = 32
ST = FourierBasis(N, np.float, plan=True)
u = TrialFunction(ST)
v = TestFunction(ST)
X = ST.mesh(N)

# Get f and exact solution on quad points 
fj = np.array([fe.subs(x, j) for j in X], dtype=np.float)
uj = np.array([ue.subs(x, i) for i in X], dtype=np.float)

# Assemble right and left hand sides
f_hat = inner(v, fj)
A = inner(v, div(grad(u)))

# Solve Poisson equation
u_hat = A.solve(f_hat)

# Transfer solution back to real space
uq = ST.backward(u_hat)
assert np.allclose(uj, uq)
\end{python}
\caption{\inpyth{shenfun} implementation of the periodic 1D Poisson problem.}
\label{fig:poisson1D}
\end{figure}

Naturally, this simple problem could be solved easier with a Fourier collocation instead, and  a simple pure 1D Fourier problem does not illuminate the true advantages of  \inpyth{shenfun}, that only will become evident when we look at higher dimensional problems with tensor product spaces. To solve with collocation, we could simply do
\vskip 1ex
\noindent
\begin{minipage}{\columnwidth}
\begin{python}
# Transform right hand side
f_hat = ST.forward(fj)

# Wavenumers
k = ST.wavenumbers(N)
k[0] = 1

# Solve Poisson equation (solution in f_hat)
f_hat /= k**2
\end{python}
\end{minipage}
Note that \inpyth{ST} methods \inpyth{forward/backward} correspond to forward and inverse discrete Fourier transforms. Furthermore, since the input data \inpyth{fj} is of type float (not complex), the transforms make use of the symmetry of the Fourier transform of real data, that $\hat{u}_k = \overline{\hat{u}}_{N-k}$, and that $\bs{k}=0,1,\ldots, N/2$ (index set computed as \inpyth{k = ST.wavenumbers(N)}).

\subsubsection{Dirichlet boundary conditions}
\label{sec:dirichletpoisson}
If the Poisson equation is subject to Dirichlet boundary conditions on the edge of the domain $\Omega \in [-1, 1]$, then a natural choice is to use Chebyshev or Legendre polynomials. Two test functions that strongly fixes the boundary condition $u(\pm 1)=0$ are
\begin{equation}
v_l(x) = T_l(x) - T_{l+2}(x),
\end{equation}
where $T_l(x)$ is the l'th order Chebyshev polynomial of the first kind, or
\begin{equation}
v_l(x) = L_l(x) - L_{l+2}(x),
\label{eq:shen_legendre_basis}
\end{equation}
where $L_l(x)$ is the l'th order Legendre polynomial. The test functions give rise to functionspaces
\begin{align}
V^C &= \text{span}\{T_l-T_{l+2}, l \in \bs{l}^D\}, \label{eq:chebdirichlet} \\
V^L &= \text{span}\{L_l-L_{l+2}, k \in \bs{l}^D\}, \label{eq:legdirichlet}
\end{align}
where
\begin{equation}
\boldsymbol{l}^D = 0, 1, \ldots, N-3.
\end{equation}
The computational mesh and associated weights will be decided by the chosen quadrature rule. Here we will go for Gauss quadrature, which leads to the following points and weights for the Chebyshev basis
\begin{align}
x_j^C &= \cos \left( \frac{2j+1}{2N}\pi \right) \quad &j=0,1,\ldots, N-1, \\
w_j^C &= \frac{\pi}{N},
\end{align}
and
\begin{align}
x_j^L &= \text{ zeros of }L_{N}(x) \quad &j=0,1,\ldots, N-1, \\
w_j^L &= \frac{2}{(1-x_j^2)[L'_{N}(x_j)]^2} \quad &j=0,1,\ldots, N-1,
\end{align}
for the Legendre basis.

We now follow the same procedure as in Sec.~\ref{sec:fourierpoisson} and solve Eq. (\ref{eq:poisson}) with the spectral Galerkin method. Consider first the Chebyshev basis and find $u \in V^C$ , such that
\begin{equation}
(-u'', v)_w^N = (f, v)_w^N , \quad \forall \, v \in V^C.
\end{equation}
We insert for $v=v_m$ and $u=\displaystyle \sum_{l\in \bs{l}^D} \hat{u}_l v_l$ and obtain
\begin{align}
-(\sum_{l\in \bs{l}^D} \hat{u}_l v_l'', v_m)_w^N &= (f, v_m)_w^N  &m \in \bs{l}^D,\\
-(v_l'', v_m)_w^N \hat{u}_l &= (f, v_m)_w^N & m \in \bs{l}^D, \label{eq:cheb_poisson}
\end{align}
where summation on repeated indices is implied. In Eq.~(\ref{eq:cheb_poisson}) $A_{ml} =(v_l'', v_m)_w^N$ are the components of a sparse stiffness matrix, and we will use matrix notation $\bs{A} = \{A_{ml}\}_{m,l \in \bs{l}^D \times \bs{l}^D}$ to simplify. The right hand side can similarily be assembled to a vector with components $\tilde{f}_m = (f, v_m)_w^N$ such that $\bs{\tilde{f}} = \{\tilde{f}_m\}_{m\in \bs{l}^D} $. Note that a tilde is used since this is not a complete transform. We can now solve for the unknown $\bs{\hat{u}} = \{\hat{u}_l\}_{l\in \bs{l}^D}$ vector
\begin{align}
-\bs{A} \bs{\hat{u}} &= \bs{\tilde{f}}, \\
   \bs{\hat{u}} &= -\bs{A}^{-1} \bs{\tilde{f}}.
\end{align}
Note that the matrix $\bs{A}$ is a special kind of upper triangular matrix, and that the solution can be obtained very efficiently in approximately $4 N$ arithmetic operations. 

To get the solution back and forth between real and spectral space we require a transformation pair similar to the Fourier transforms. We do this by projection. Start with
\begin{equation}
u(\bs{x}) = \sum_{l\in \bs{l}^D} \hat{u}_l v_l(\bs{x})
\end{equation}
and take the inner product with $v_m$
\begin{equation}
(u, v_m)_w^N  = (\sum_{l\in \bs{l}^D} \hat{u}_l v_l, v_m)_w^N.
\label{eq:projection}
\end{equation}
Introducing now the mass matrix $B_{ml} = (v_l, v_m)_w^N$ and the \emph{Shen} forward inner product $\mathcal{S}_m(u) = (u, v_m)_w^N$, Eq. (\ref{eq:projection})  is rewritten as
\begin{align}
\mathcal{S}_m(u) &= B_{ml} \hat{u}_l, \\
\bs{\hat{u}}  =& \bs{B}^{-1} \mathcal{S}(\bs{u}) , \\
\bs{\hat{u}}  =& \mathcal{T}(\bs{u}) ,
\end{align}
where $\mathcal{T}(\bs{u})$ represents a forward transform of $\bs{u}$. Note that $\mathcal{S}$ is introduced since the inner product $(u, v_m)_w^N$ may, just like the inner product with the Fourier basis, be computed fast, with $\mathcal{O}(N \log N)$ operations. And to this end, we need to make use of a discrete cosine transform (DCT), instead of the Fourier transform. The details are left out from this paper, though.

A simple Poisson problem with analytical solution $\sin(\pi x)(1-x^2)$ is implemented in Figure~\ref{fig:poisson1D_dirichlet}, where we also verify that the correct solution is obtained.
\begin{figure}[h!]
\begin{python}
from shenfun.chebyshev.bases import ShenDirichletBasis

# Use sympy to compute a rhs, given an analytical solution
ue = sin(np.pi*x)*(1-x**2)
fe = ue.diff(x, 2)

# Lambdify for faster evaluation
ul = lambdify(x, ue, 'numpy')
fl = lambdify(x, fe, 'numpy')

N = 32
SD = ShenDirichletBasis(N, plan=True)
X = SD.mesh(N)
u = TrialFunction(SD)
v = TestFunction(SD)
fj = fl(X)

# Compute right hand side of Poisson equation
f_hat = inner(v, fj)

# Get left hand side of Poisson equation and solve
A = inner(v, div(grad(u)))
f_hat = A.solve(f_hat)
uj = SD.backward(f_hat)

# Compare with analytical solution
ue = ul(X)
assert np.allclose(uj, ue)
\end{python}
\caption{\inpyth{shenfun} implementation of the 1D Poisson problem with Dirichlet boundary conditions.}
\label{fig:poisson1D_dirichlet}
\end{figure}
Note that the inner product \inpyth{f_hat = inner(v, fj)} is computed under the hood using the fast DCT.  The inverse transform \inpyth{uj = SD.backward(f_hat)} is also computed using a fast DCT, and we use the notation
\begin{align}
u(x_j) &= \sum_{l\in \bs{l}^D} \hat{u}_l v_l(x_j) &j=0,1,\ldots, N-1, \notag \\
\bs{u} &= \mathcal{S}^{-1}(\bs{\hat{u}}). \label{eq:fast_shen}
\end{align}
To implement the same problem with the Legendre basis (\ref{eq:shen_legendre_basis}), all that is needed to change is the first line in Fig~\ref{fig:poisson1D_dirichlet} to \inpyth{from shenfun.legendre.bases import ShenDirichletBasis}. Everything else is exactly the same. However, a fast inner product, like in (\ref{eq:fast_shen}), is only implemented for the Chebyshev basis, since there are no known $\mathcal{O}(N \log N)$ algorithms for the Legendre basis, and the Legendre basis thus uses straight forward $\mathcal{O}(N^2)$ algorithms for its transforms.

\section{TENSOR PRODUCT SPACES}
\label{sec:tensorproductspaces}
Now that we know how to solve problems in one dimension, it is time to move on to more challenging tasks. Consider again the Poisson equation, but now in possibly more than one dimension
\begin{align}
 -\nabla^2 u(\bs{x}) &= f(\bs{x}) & \bs{x} \in \Omega.
\end{align}
Lets first consider 2 dimensions, with Dirichlet boundary conditions in the first direction and with periodicity in the second. Let $\Omega$ be the domain $ [-1, 1] \times [0, 2\pi]$, and $W(x,y) = V^C(x) \times V^p(y)$ be the tensor product function space. We can solve this problem for some suitable function $f(\bs{x})$ in \inpyth{shenfun} by constructing a few more classes than were required in 1D
\begin{python}
from shenfun import Function, TensorProductSpace
from mpi4py import MPI
\end{python}
Now the \inpyth{TensorProductSpace} class is used to construct $W$, whereas \inpyth{Function} is a subclass of \inpyth{numpy.ndarray} used to hold solution arrays. The MPI communicator, on the other hand, is used for distributing the tensor product grids on a given number of processes
\vskip 1ex
\noindent
\begin{minipage}{\columnwidth}
\begin{python}
comm = MPI.COMM_WORLD
N = (32, 33)

K0 = ShenDirichletBasis(N[0])
K1 = FourierBasis(N[1], dtype=np.float)
W = TensorProductSpace(comm, (K0, K1))

# Alternatively, switch order for periodic in first direction instead
# W = TensorProductSpace(comm, (K1, K0), axes=(1, 0))
\end{python}
\end{minipage}
Under the hood, within the \inpyth{TensorProductSpace} class, the mesh is distributed, both in real, physical space, and in spectral space. In the real space the mesh is distributed along the first index, whereas in spectral space the wavenumbermesh is distributed along the second dimension. This is the default behaviour of \inpyth{TensorProductSpace}. However, the distribution may also be configured specifically by the user, e.g., as shown in the commented out text, where the Dirichlet basis is found along the second axis. In this case the order of the axes to transform over has been flipped, such that in spectral space the data is distributed along the first dimension and aligned in the second. This is required for solving the linear algebra system that arises for the Dirichlet basis. The arrays created using \inpyth{Function} are distributed, and no further attention to MPI is required. However, note that arrays may have different type and shape in real space and in spectral space. For this reason \inpyth{Function} has a keyword argument \inpyth{forward_output}, that is used as \inpyth{w_hat = Function(W, forward_output=True)} to create an array consistent with the output of \inpyth{W.forward} (solution in spectral space), and \inpyth{w = Function(W, forward_output=False)} to create an array consistent with the input (solution in real space). Furthermore, \inpyth{uh = np.zeros_like(w_hat); w_hat = Function(W, buffer=uh)} can be used to wrap a \inpyth{Function} instance around a regular Numpy array \inpyth{uh}. Note that \inpyth{uh} and \inpyth{w_hat} now will share the same data, and modifying one will naturally modify also the other. 

The solution of a complete Poisson problem in 2D is shown in Figure~\ref{fig:poisson2D}. Very similar code is required to solve the Poisson problem with the Legendre basis. The main difference is that for Legendre it is natural to integrate the weak form by parts and use
\begin{python}
matrices = inner(grad(v), grad(u))
\end{python}
\begin{figure}[h!]
\begin{python}
from shenfun.chebyshev.la import Helmholtz as Solver

# Create a solution that satisfies boundary conditions
x, y = symbols("x,y")
ue = (cos(4*y) + sin(2*x))*(1-x**2)
fe = ue.diff(x, 2) + ue.diff(y, 2)

# Lambdify for faster evaluation
ul = lambdify((x, y), ue, 'numpy')
fl = lambdify((x, y), fe, 'numpy')

X = T.local_mesh(True)
u = TrialFunction(T)
v = TestFunction(T)

# Get f on quad points
fj = fl(X[0], X[1])

# Compute right hand side of Poisson equation
f_hat = inner(v, fj)

# Get left hand side of Poisson equation
matrices = inner(v, div(grad(u)))

# Create Helmholtz linear algebra solver
H = Solver(**matrices)

# Solve and transform to real space
u_hat = Function(T)           # Solution spectral space
u_hat = H(u_hat, f_hat)       # Solve
u = T.backward(u_hat)
\end{python}
\caption{Solution of Poisson equation in 2D using Dirichlet boundary conditions in the $x$-direction, and periodic boundaries in the $y$-direction.}
\label{fig:poisson2D}
\end{figure}

The test functions and function spaces require a bit more attention. Test functions for space $W(x, y)=V^C(x) \times V^p(y)$ are given as
\begin{equation}
\phi_{\ts{k}}(x, y) = v_l(x) e^{imy},
\end{equation}
which introduces the sans serif tensor product wavenumber mesh $\ts{k} = \bs{l}^D \times \bs{l}$
\begin{equation}
 \ts{k} = \{ (l, m) | l \in \bs{l}^D \text{ and } m \in \bs{l}\}.
\end{equation}
Similarly there is a tensor product grid $\ts{x} = \bs{x} \times \bs{y}$, where $\bs{y} = \{y_k\}_{k=0}^{M-1} = 2 \pi k /M$
\begin{equation}
 \ts{x} = \{ (x_j, y_k) | j=0,1,\ldots, N-1 \text{ and } k=0,1,\ldots, M-1\}.
\end{equation}
Note that for computing on the tensor product grids using Numpy arrays with vectorization, the mesh and wavenumber components need to be represented as 2D arrays. As such we create
\begin{equation}
\bs{\textsf{x}} = (\bs{x}, \bs{y}) = \Big(\{x_i\}_{i=0}^{N-1} \times I^M,  I^N \times \{y_j\}_{j=0}^{M-1} \Big),
\end{equation}
where $I^N$ is an N-length vector of ones. Similarly
\begin{equation}
\bs{\textsf{k}} = (\bs{l}, \bs{m}) = \Big(\{ l \}_{l=0}^{N-1} \times I^M,  I^N \times \{ m \}_{m=0}^{M/2} \Big). 
\end{equation}
Such tensor product grids can be very efficiently stored with Numpy arrays, using no more space than the two vectors used to create them. The key to this efficiency is broadcasting. We store $\ts{k}$ as a list of two numpy arrays, $\bs{l}$ and $\bs{m}$, corresponding to the two 1D wavenumber meshes $\{ l \}_{l=0}^{N-1}$ and $\{ m \}_{m=0}^{M/2}$. 
However, $\bs{l}$ and $\bs{m}$ are now stored as 2D arrays of shape $(N, 1)$ and $(1, M/2+1)$, respectively. And broadcasting takes care of the additional dimension, such that the two arrays work just like if they were stored as $(N, M/2+1)$ arrays. We can look up $\bs{l}(l, m)$, just like a regular $(N, M/2+1)$ array, but the storage required is still only one single vector. 
The same goes for $\ts{x}$, which is stored as a list of two arrays $\bs{x}$, $\bs{y}$ of shape $(N, 1)$ and $(1, M)$ respectively. This extends straightforward to even higher dimensions. 

Assembling a weak form like $(v, \nabla^2 u)_w^N$ leads to two non-diagonal matrices, both the stiffness and mass matrix, since it expands like
\begin{equation}
(v, \nabla^2 u)_w^N = \left(v, \frac{\partial^2 u}{\partial x^2} + \frac{\partial^2 u}{\partial y^2} \right)_w^N.
\end{equation}
Inserting for test function $v = \phi_{\ts{k}} (= \phi_{l, m} =v_l(x) e^{imy})$ and trial function $u = \sum_{(q,r)\in \ts{k}} \hat{u}_{q, r} \phi_{q,r}$, we obtain
\begin{align}
 (v, \nabla^2 u)_w^N &= \left(\phi_{l, m}, \frac{\partial^2}{\partial x^2} \sum_{(q, r) \in \ts{k}} \hat{u}_{q, r} \phi_{q, r} + \frac{\partial^2}{\partial y^2} \sum_{(q,r) \in \ts{k}} \hat{u}_{q, r} \phi_{q, r} \right)_w^N, \\
 &= 2\pi \left(\sum_{(q, r) \in \ts{k}} A_{lq} \delta_{rm} \hat{u}_{q,r} -  \sum_{(q, r) \in \ts{k}} {r}^2  B_{lq} \delta_{rm} \hat{u}_{q,r}\right), \\
 &= 2\pi \left(\sum_{q\in \bs{l}^D} A_{lq} \hat{u}_{q,m} - {m}^2 \sum_{q\in \bs{l}^D}  B_{lq} \hat{u}_{q,m}\right) \quad \forall (l, m) \in \bs{l}^D \times \bs{l}. \label{eq:laplace}
\end{align}
As can be seen from Eq.~(\ref{eq:laplace}), the linear system of equations is set up to act along the Dirichlet direction, whereas for the periodic direction the matrices are diagonal and no additional work is required. The system of equations correspond to a series of 1D Helmholtz problems, that need to be solved once for each $m \in \bs{l}$. This is what goes on under the hood with the Helmholtz solver imported through \inpyth{from shenfun.chebyshev.la import Helmholtz as Solver}.

The right hand side of the Poisson problem is computed as
\begin{align}
(v, f)_w^N &= 2\pi \underbrace{\sum_{j}\underbrace{\frac{1}{N} \sum_{k} f(x_j, y_k) e^{imy_k} }_{\mathcal{F}_m} v_l(x_j)   w_j}_{\mathcal{S}_l} \quad \forall (l, m) \in \bs{l}^D \times \bs{l}, \notag \\
  &= 2\pi \mathcal{S}(f) = 2 \pi \mathcal{S}_l(\mathcal{F}_m(f)).
\end{align}

The \inpyth{TensorProductSpace} class can take any number of Fourier bases. A 3 dimensional tensor product space can be created as
\begin{python}
N = (32, 33, 34)
K0 = ShenDirichletBasis(N[0])
K1 = C2CBasis(N[1])
K2 = R2CBasis(N[2])
W = TensorProductSpace(comm, (K0, K1, K2))
\end{python}
Here the default behaviour of \inpyth{TensorProductSpace} is to distribute the first 2 indices in real space using two subcommunicators, with a decomposition often referred to as \emph{pencil} decomposition. In spectral space the last two indices will be distributed. For example, using 4 CPUs, a subprocessor mesh of size $2 \times 2$ will be created, and 2 subprocessors share the first index and the other two share the second index.  If the program is run with 3 processors, then only the first index will be distributed and the subprocessormesh will be $3 \times 1$. It is also possible to configure \inpyth{TensorProductSpace} to run with 4 CPUs and a $4 \times 1$ subprocessormesh, or 40,000 CPUs with a $200 \times 200$ processormesh. The latter requires that the mesh is big enough, though, but otherwise it is just a matter of acquiring computing power. The biggest simulations tested thus far used 64,000 CPUs. 

Solving a biharmonic problem is just as easy as the Poisson problem. Consider the fourth order biharmonic PDE in 3-dimensional space
\begin{align}
\nabla^4 u(\bs{x}) &= f(\bs{x}), \quad \bs{x} \in \Omega \\
 u(x=\pm1, y, z) &= \frac{\partial u}{\partial x} (x=\pm 1, y, z) = 0 \\
 u(x, y+2\pi, z) &= u(x, y, z), \\
 u(x, y, z+2\pi) &= u(x, y, z). 
\end{align}
that is periodic in $y-$ and $z-$directions and with clamped boundary conditions at $x=\pm 1$. The problem may be solved using either one of these two bases:
\begin{align}
V^C &= \text{span}\{T_l - \frac{2(l+2)}{l+3}T_{l+2} + \frac{l+1}{l+3}T_{l+4} , l \in \bs{l}^B\}, \label{eq:chebbiharmonic} \\
V^L &= \text{span}\{L_l - \frac{2(2k+5)}{2k+7}L_{l+2} + \frac{2k+3}{2k+7}, l \in \bs{l}^B\}, \label{eq:legbiharmonic}
\end{align}
where $\bs{l}^B = 0, 1, \ldots, N-5$. A tensor product space may be constructed as $W(x,y,z) = V^C(x) \times V^p(y) \times V^p(z)$, and the variational problem 
\begin{equation}
(v, \nabla^4 u)^N_w = (v, f)^N_w,
\end{equation}
where $u$ and $v$ are trial and test functions in $W$, may be implemented in \inpyth{shenfun} as shown below
\vskip 1ex
\noindent
\begin{minipage}{\columnwidth}
\begin{python}
from shenfun.chebyshev.bases import ShenBiharmonicBasis
from shenfun.chebyshev.la import Biharmonic as Solver

N = (32, 33, 34)
K0 = ShenBiharmonicBasis(N[0])
K1 = C2CBasis(N[1])
K2 = R2CBasis(N[2])
W = TensorProductSpace(comm, (K0, K1, K2))
u = TrialFunction(W)
v = TestFunction(W)
matrices = inner(v, div(grad(div(grad(u)))))
f_hat = inner(v, fj)  # Some right hand side
# or for Legendre:
# matrices = inner(div(grad(v)), div(grad(u)))
B = Solver(**matrices)

# Solve and transform to real space
u_hat = Function(T)           # Solution spectral space
u_hat = B(u_hat, f_hat)       # Solve
u = T.backward(u_hat)
\end{python}
\end{minipage}

\section{OTHER FUNCTIONALITY OF \inpyth{SHENFUN}}
\label{sec:extended}
In addition to the \inpyth{div} and \inpyth{grad} operators, there is \inpyth{Dx} for a partial derivative
\vskip 1ex
\noindent
\begin{minipage}{\columnwidth}
\begin{python}
from shenfun import Dx
v = TestFunction(W)
du = Dx(v, 0, 1)
\end{python}
\end{minipage}
where the first argument is the basis function, the second (integer) is the axis to take the derivative over, and the third (integer) is the number of derivatives, e.g.,
\begin{equation}
\frac{\partial^2 v}{\partial y^2} = \text{Dx(v, 1, 2)}. \notag
\end{equation}
The operator can be nested. To compute $\frac{\partial^2 u}{\partial x  \partial y} $ one may do
\vskip 1ex
\noindent
\begin{minipage}{\columnwidth}
\begin{python}
v = TestFunction(W)
du = Dx(Dx(v, 0, 1), 1, 1)
\end{python}
\end{minipage}
The operators work on \inpyth{TestFunctions, TrialFunctions} or \inpyth{Functions}, where only the last actually contain any data, because a \inpyth{Function} is used to store the solution. Once a solution has been found, one may also manipulate it further using \inpyth{project} in combination with operators on \inpyth{Function}s. For example, to compute $\partial u / \partial x$ of the solution to the biharmonic problem, one can do
\vskip 1ex
\noindent
\begin{minipage}{\columnwidth}
\begin{python}
u = T.backward(u_hat)  # The original solution on space T
K0 = Basis(N[0])
W0 = TensorProductSpace(comm, (K0, K1, K2))
du_hat = project(Dx(u, 0, 1), W0, uh_hat=u_hat)
du = Function(W0)
du = W0.backward(du_hat, du)
\end{python}
\end{minipage}
Note that we are here using a regular Chebyshev space instead of the biharmonic, to avoid enforcing erroneous boundary conditions on the solution. We could in this case also, with advantage, have chosen a Dirichlet space, since the derivative of the biharmonic problem is known to be zero on the edges of the domain (at $x=\pm 1$).

All problems considered thus far have been scalar valued. With \inpyth{shenfun} there is also some functionality for working with vector equations. To this end, there is a class called \inpyth{VectorTensorProductSpace}, and there is an additional operator, \inpyth{curl}, that can only be used on vectors:
\vskip 1ex
\noindent
\begin{minipage}{\columnwidth}
\begin{python}
from shenfun import VectorTensorProductSpace, curl
T = TensorProductSpace(comm, (K0, K1, K2))
Tk = VectorTensorProductSpace([T, T, T])
v = TestFunction(Tk)
u_ = Function(Tk, False)
u_[:] = np.random.random(u_.shape)
u_hat = Tk.forward(u_)
w_hat = inner(v, curl(u_), uh_hat=u_hat)
\end{python}
\end{minipage}
Vector equations have very similar form as scalar equations, but at the moment of writing the different equation components cannot be implicitly coupled.

\section{GINZBURG-LANDAU EQUATION}
\label{sec:ginzburg}
We end this paper with the implementation of the complex Ginzburg-Landau equation, which is a  nonlinear time dependent reaction-diffusion problem. The equation to solve is 
\begin{equation}
\frac{\partial u}{\partial t} = \nabla^2u + u - (1 + 1.5i)u |u|^2,
\end{equation}
for the doubly periodic domain $\Omega = [-50, 50]\times [-50, 50]$ and  $t \in [0, T]$. The initial condition is chosen as one of the following
\begin{align}
u^0(\boldsymbol{x}, 0) &= (ix + y) \exp {-0.03 (x^2 + y^2)} \label{eq:initial_0}, \\
u^1(\boldsymbol{x}, 0) &= (x + y) \exp {-0.03 (x^2 + y^2)} \label{eq:initial_1}.
\end{align}
This problem is solved with the spectral Galerkin method using Fourier bases in both directions, and a tensor product space $W(x,y)=V^p(x) \times V^p(y)$, where $V^p$ is defined as in Sec~\ref{sec:fourierpoisson}, but here mapping the computational domain $[-50, 50]$ to $[0, 2\pi]$. Considering only the spatial discretization, the variational problem becomes: find $u(x, y)$ in $W$, such that
\begin{equation}
\frac{\partial }{\partial t} (v, u)^N = (v, \nabla^2u)^N + (v, u - (1 + 1.5i)u |u|^2)^N \quad \text{for all} \quad v \in W, \label{eq:Ginz_var}
\end{equation}
and we integrate the equations forward in time using an explicit, fourth order Runge-Kutta method, that only requires as input a function that returns the right hand side of (\ref{eq:Ginz_var}). Note that all matrices involved with the Fourier method are diagonal, so there is no need for linear algebra solvers, and the left hand side inner product equals $(2 \pi)^2 \bs{\hat{u}}$.

The initial condition is created using \inpyth{Sympy}
\vskip 1ex
\noindent
\begin{minipage}{\columnwidth}
\begin{python}
from sympy import symbols, exp, lambdify
x, y = symbols("x,y")
#ue = (1j*x + y)*exp(-0.03*(x**2+y**2))
ue = (x + y)*exp(-0.03*(x**2+y**2))
ul = lambdify((x, y), ue, 'numpy')
\end{python}
\end{minipage}
We create a regular tensor product space, choosing the \inpyth{fourier.bases.C2CBasis} for both directions if the initial condition is complex (\ref{eq:initial_0}), whereas we may choose \inpyth{R2CBasis} if the initial condition is real (\ref{eq:initial_1}). Since we are solving a nonlinear equation, the additional issue of aliasing should be considered. Aliasing errors may be handled with different methods, but here we will use the so-called 3/2-rule, that requires padded transforms. We create a tensor product space \inpyth{Tp} for padded transforms, using the \inpyth{padding_factor=3/2} keyword below. Furthermore, some solution arrays, test and trial functions are also declared.
\vskip 1ex
\noindent
\begin{minipage}{\columnwidth}
\begin{python}
# Size of discretization
N = (201, 201)

# Create tensor product space
K0 = C2CBasis(N[0], domain=(-50., 50.))
K1 = C2CBasis(N[1], domain=(-50., 50.))
T = TensorProductSpace(comm, (K0, K1))

Kp0 = C2CBasis(N[0], domain=(-50., 50.), padding_factor=1.5)
Kp1 = C2CBasis(N[1], domain=(-50., 50.), padding_factor=1.5)
Tp = TensorProductSpace(comm, (Kp0, Kp1))

u = TrialFunction(T)
v = TestFunction(T)
X = T.local_mesh(True)
U = Function(T, False)         # Solution
U_hat = Function(T)            # Solution spectral space
Up = Function(Tp, False)       # Padded solution for nonlinear term
dU_hat = Function(T)           # right hand side
#initialize
U[:] = ul(*X)
U_hat = T.forward(U, U_hat)
\end{python}
\end{minipage}
Note that \inpyth{Tp} can be used exactly like \inpyth{T}, but that a backward transform creates an output that is 3/2 as large in each direction. So a $(100, 100)$ mesh results in a $(150, 150)$ output from a backwards transform. This transform is performed by creating a 3/2 times larger padded array in spectral space $\hat{u}^p_{\textsf{k}^p}$, where $\textsf{k}^p = \boldsymbol{l}^p \times \boldsymbol{l}^p $ and
\begin{equation}
\boldsymbol{l}^{p} = -3N/4, -3N/4+1, \ldots, 3N/4-1.
\end{equation}
We then set $\hat{u}^p_{\textsf{k}} = \hat{u}_{\textsf{k}}$ for $\textsf{k} \in \boldsymbol{l} \times \boldsymbol{l}$, and for the remaining high frequencies $\hat{u}^p_{\textsf{k}}$ is set to 0.

\begin{figure}[t!]
    \centering
    \subfigure[t=16, initial condition (\ref{eq:initial_0}).]{
        \centering
        \includegraphics[scale=0.4]{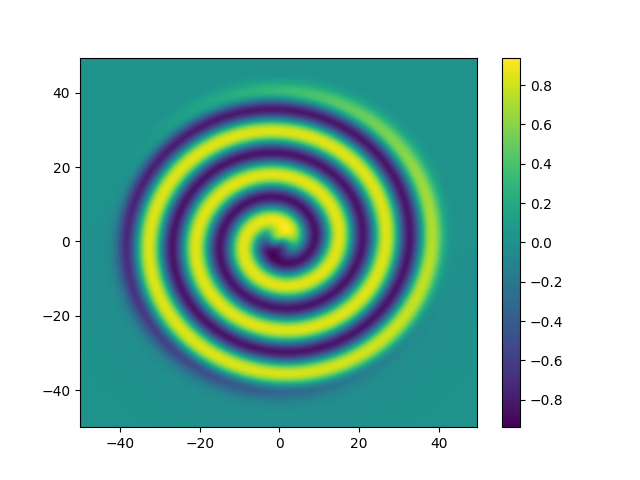}
        \label{subfig0}
    }%
    ~ 
    \subfigure[t=96, initial condition (\ref{eq:initial_0}).]{
        \centering
        \includegraphics[scale=0.4]{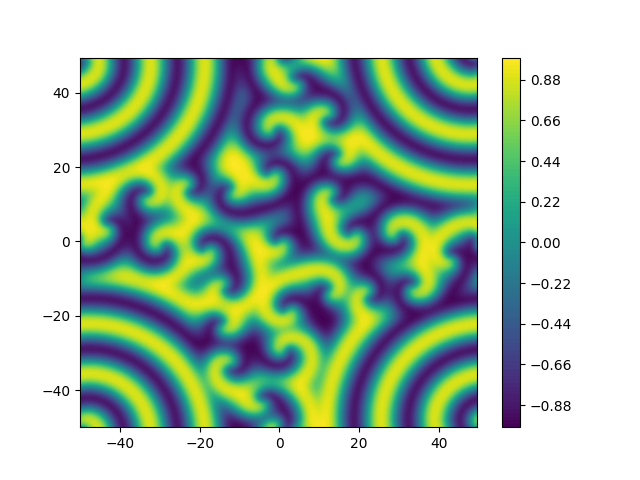}
        \label{subfig1}
    }

    \subfigure[t=16, initial condition (\ref{eq:initial_1}).]{
        \centering
        \includegraphics[scale=0.4]{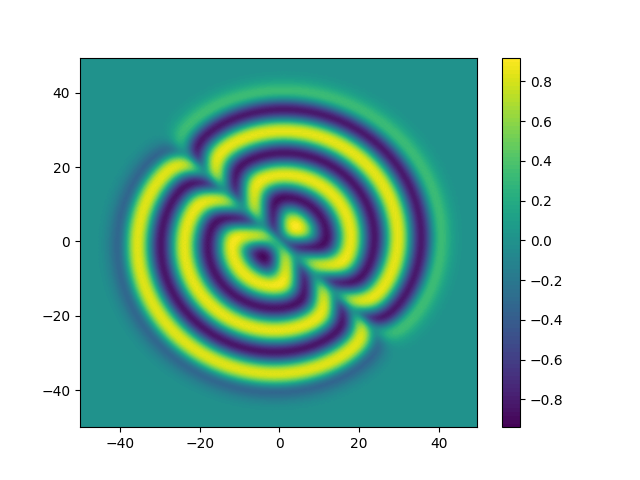}
        \label{subfig2}
    }%
    ~ 
    \subfigure[t=96, initial condition (\ref{eq:initial_1}).]{
        \centering
        \includegraphics[scale=0.4]{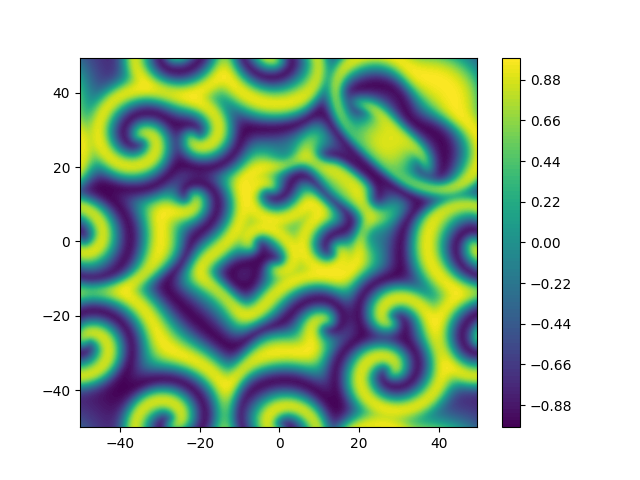}
        \label{subfig3}
    }

    \caption{Real part of solution to Ginzburg-Landau problem with real initial condition (\ref{eq:initial_0}), at two different times in \subref{subfig0} and \subref{subfig1}, and for initial condition (\ref{eq:initial_1}) in \subref{subfig2} and \subref{subfig3}. The mesh is of size $201 \time 201$ in both cases.}
    \label{fig:GL}
\end{figure}

We will solve the equation with a fourth order Runge-Kutta integrator. To this end we need to declare some work arrays to hold intermediate solutions, and a function for the right hand side of Eq.~(\ref{eq:Ginz_var})
\vskip 1ex
\noindent
\begin{minipage}{\columnwidth}
\begin{python}
U_hat0 = Function(T)
U_hat1 = Function(T)
w0 = Function(T)
a = [1./6., 1./3., 1./3., 1./6.]         # Runge-Kutta parameter
b = [0.5, 0.5, 1.]                       # Runge-Kutta parameter
def compute_rhs(rhs, u_hat, U, Up, T, Tp, w0):
    rhs.fill(0)
    U = T.backward(u_hat, U)
    rhs = inner(v, div(grad(U)), output_array=rhs, uh_hat=u_hat)
    rhs += inner(v, U, output_array=w0, uh_hat=u_hat)
    rhs /= (2*np.pi)**2  # (2pi)**2 represents scaling with inner(u, v)
    Up = Tp.backward(u_hat, Up)
    rhs -= Tp.forward((1+1.5j)*Up*abs(Up)**2, w0)
    return rhs
\end{python}
\end{minipage}
Note the close similarity with (\ref{eq:Ginz_var}) and the usage of the padded transform for the nonlinear term.
Finally, the Runge-Kutta method is implemented as
\vskip 1ex
\noindent
\begin{minipage}{\columnwidth}
\begin{python}
t = 0.0
dt = 0.025
end_time = 96.0
tstep = 0
while t < end_time-1e-8:
    t += dt
    tstep += 1
    U_hat1[:] = U_hat0[:] = U_hat
    for rk in range(4):
        dU_hat = compute_rhs(dU_hat, U_hat, U, Up, T, Tp, w0)
        if rk < 3:
            U_hat[:] = U_hat0 + b[rk]*dt*dU_hat
        U_hat1 += a[rk]*dt*dU_hat
    U_hat[:] = U_hat1
\end{python}
\end{minipage}
The code that is described here will run in parallel for up to a maximum of $\text{min}(N[0], N[1])$ processors. But, being a 2D problem, a single processor is sufficient to solve the problem in reasonable time. The real part of $u(\boldsymbol{x}, t)$ is shown in Figure \ref{fig:GL} for times $t=16$ and $t=96$. The results from using the complex initial condition (\ref{eq:initial_0}) are shown in subfigures (a) and (b), whereas the results starting from the real initial condition in (\ref{eq:initial_1}) are shown in (c) and (d). There are apparently good agreements with figures published from using \inpyth{chebfun} on \emph{www.chebfun.org/examples/pde/GinzburgLandau.html}. In particular, figures (a) and (b) are identical by the eye norm. One interesting feature, though, is seen in subfigure (d), where the results can be seen to have preserved symmetry, as they should. This symmetry is lost with chebfun, as commented in the referenced webpage. An asymmetric solution is also obtained with \inpyth{shenfun} if no de-aliasing is applied. However, the simulations are very sensitive to roundoff, and it has also been observed that a de-aliased solution using \inpyth{shenfun} may loose symmetry simply if a different FFT algorithm is chosen on runtime by FFTW.

\section{CONCLUSIONS}
In this paper, the Python module \inpyth{shenfun} has been described. Within this module there are tools that greatly simplify the implementation of the spectral Galerkin method for tensor product grids, and parallel solvers may be written with ease and comfort.  \inpyth{Shenfun} provides a FEniCS like interface to the spectral Galerkin method, where equations are cast on a weak form, and where the required script-like coding remains very similar to the mathematics. We have verified and shown implementations for simple Poisson or biharmonic problems, as well as the nonlinear complex Ginzburg-Landau equation. On a final note, it  should be mentioned that these tools have also been used to implement various Navier Stokes solvers within the \inpyth{spectralDNS} project (github.com/spectralDNS), that has run on the Shaheen II supercomputer at KAUST, on meshes of size up to $2048^3$.

\section*{ACKNOWLEDGEMENTS}
This research is a part of the 4DSpace Strategic Research Initiative at the University of Oslo. 

\bibliographystyle{model1-num-names}
\bibliography{mekit17.bib}

\end{document}